\documentclass[pre,superscriptaddress]{revtex4}
\usepackage{epsfig}
\usepackage{latexsym}
\usepackage{amsmath}
\begin {document}
\title {Some recent developments in models with absorbing states}
\author{Michel Droz}
\affiliation{Department of Physics, University of Geneva, CH 1211
Geneva 4, Switzerland}
\author{Adam Lipowski}
\affiliation{Department of Physics, University of Geneva, CH 1211
Geneva 4, Switzerland}
\affiliation{Faculty of Physics, A.~Mickiewicz University,
61-614 Pozna\'{n}, Poland}
\pacs{}
\begin {abstract}
We describe some of the recent results obtained for models with absorbing 
states.
First, we present the nonequilibrium absorbing-state Potts model and discuss 
some of the factors that might affect the critical behaviour of such models.
In particular we show that in two dimensions the further neighbour interactions 
might split the voter critical point into two critical points.
We also describe some of the results obtained in the context of synchronization 
of chaotic dynamical systems.
Moreover, we discuss the relation of the synchronization transition with some 
interfacial models.
\end{abstract}
\maketitle
\section{Introduction}

Nonequilibrium statistical mechanics is nowaday a very active research 
field~\cite{MARRODICKMAN} and there are several reasons for that. 
First, the most interesting phenomena in 
Nature take place out of equilibrium. The best example is provided by living 
matter. More generally,  systems which are open (traversed by fluxes of energy, 
entropy or matter) may reach stationary states which cannot be described by 
equilibrium statistical physics. Second, the properties of systems in 
equilibrium are by now rather well understood as equilibrium statistical 
physics is a well established theory. The puzzling problem of universal 
behavior observed in the vicinity of second order phase transitions is 
beautifully
explained by the renormalization group approach~\cite{RG}. It is thus natural 
to try to extend our understanding of equilibrium systems to nonequilibrium 
ones.  

Indeed, the situation is not so clear for nonequilibrium statistical
mechanics for which no general theory has been developed yet. This is
particularly true for the case of nonequilibrium phase transitions
where, according to the values of some control parameters, the system
can change, continuously or not, from one stationary state to another 
one~\cite{MARRODICKMAN}.

At the microscopic level, models for nonequilibrium phase transitions are 
usually defined in term of a master equation~\cite{GARDINER}. Most of the 
physics is contained in the transition rates. One of the key difference between 
equilibrium and nonequilibrium systems is related with the detailed-balance 
condition which is generally not obeyed in nonequilibrium. As a result, if is 
often impossible to find an analytical solution to the master equation, even 
for the stationary state. This is why a lot
of results in this field are obtained by numerical simulations.
At a coarse-grained level, the physics is often described in terms of 
a generalized Langevin equation. Unfortunately, there is no general method 
which 
allows us to perform, in a controlled way, the coarse-graining process and 
usually the determination of the form taken by the Langevin equation is based 
only on general symmetry arguments~\cite{VANK,GRIEN,MUNOZ,DICK}. This procedure 
is 
particularly ambiguous for systems with noise~\cite{HOWARD}. When 
the noise is multiplicative different interpretations of the stochastic process 
described by the master equation are possible, leading to further 
confusion~\cite{VANKB, DENBROECK}.
 
There are many examples of nonequilibrium phase transitions. One of the 
simplest 
examples is  provided by an Ising like model with competing dynamics. The 
system 
is in contact with two heat baths at different temperature, each generating a 
microscopic dynamics obeying detailed-balance (see \cite{MDONE,MDTWO,PRIVMAN} 
for more details). As a result of this competition, effective long range 
interaction develops in the system and its critical properties are similar to 
the ones of equilibrium models. Another generic way to obtain nonequilibrium 
phase transition is to
induce dynamical anisotropy. This can be done in several ways. Examples are 
given by the so-called driven lattice gases~\cite{DRIVEN}, where particles 
diffusing on a lattice
are driven by an external field oriented in a particular direction. At low 
temperature, the system exhibits a nonequilibrium phase transition and the 
ordered phase is characterized by strong anisotropy. As a result, the critical 
properties belong to a new universality class, not related with equilibrium  
phase transitions.

The question of the characterization of the possible universality
classes for nonequilibrium phase transition is thus an important one
which is much debated. The effects of violating detailed-balance on the
universal static and dynamic scaling behavior has been investigated
independently by Grinstein et al.~\cite{JAPRA} and T\"auber et 
al.~\cite{TAUBER}.
It turns out that
the standard critical dynamics universality classes are rather robust
and that detailed-balance can be effectively restored at criticality in
some cases. Nevertheless, this is not always true and a complete
characterization of the nonequilibrium universality classes is clearly
a difficult and open question.

An important type of nonequilibrium phase transition is the so-called
{\it absorbing state phase transition} that 
takes place when a system during its evolution reaches
a configuration in which it remains trapped forever.
Such a state is called an absorbing state and a given system
may have one or more such states. Absorbing states are present in
numerous systems encountered in physics, chemistry and biology. A lot
of efforts have been devoted to the study of such systems and there are
some comprehensive reviews on this subject~\cite{HAYE2000,GEZA2002}.

In the present paper we describe some recently obtained results that are not 
yet covered in these reviews and that are related with our
own research in this field.
In particular, we describe a certain $d$-dimensional model with $q$ absorbing 
states whose dynamics can be considered as a modification of the Metropolis 
algorithm for the equilibrium Potts model~\cite{POTTS1}.
We describe the role of $q$ and $d$ on the critical behaviour of this model.
We also show that some other details of the dynamics, as e.g., positivity of 
certain transition rates~\cite{POTTS2} or the range of 
interactions~\cite{POTTS3}, influence the critical behaviour.
Although its dynamics looses the detailed balance property, the model still 
bears some similarity to the equilibrium Potts model, that is also reflected in 
the overall behaviour of the model in the $(q,d)$ plane.
In a certain case, when the range of interactions in our model is increased, we 
observe that
the so-called voter transition~\cite{CHATE}, that typically occurs for $q=d=2$, 
is splitted 
into two phase transitions: first a spontaneous symmetry breaking preselects 
one of the two absorbing states and then a collapse on the preselected 
absorbing state takes place~\cite{POTTS3}.
Since our model is expressed in terms of spin operators and a Hamiltonian-like 
function, one can easily construct its various generalizations that would 
include multi-spin interactions, anisotropies or higher order symmetries 
($O(N)$).
Such models might exhibit novel properties and also enrich
the collection of relations between equilibrium and nonequilibrium systems.

Recently a class of nonequilibrium  phase transitions  is intensively studied 
in the context of synchronization~\cite{FUJISAKA}.
When for example two identical chaotic dynamical systems are
sufficiently strongly connected, they might synchronize against each other.
Since the system cannot escape from the synchronized state it is actually a 
certain absorbing state.
For spatially extended dynamical systems (e.g., coupled map lattices) the 
synchronization transition (ST) resembles some absorbing state phase 
transitions that are typical to statistical mechanics models.
Some arguments were given~\cite{PIKKURTHS} that ST should generically belong to 
the same 
universality class as a certain model of a driven interface bounded with a wall 
(BKPZ).
However, for a certain class of dynamical systems ST has a different critical 
behaviour, namely it belongs to the DP universality class~\cite{LIVI}.
Despite some attempts, theoretical understanding of the mechanism that changes 
the critical behaviour is still 
missing~\cite{LIPDROZSOS,MUNOZPASTOR,AHLERSLIVI}.
An interesting related problem is a nature of a multicritical point that 
possibly joins the critical lines of DP and BKPZ type.
There are some indications that these problems might be also related 
with some particle systems like the recently intensively studied PCPD 
model~\cite{PCPD}.

In section II we discuss the Potts model with absorbing states.
The synchronization of extended chaotic systems is discussed in section III, 
and these two sections are essentially independent.
Conclusions are presented in section IV.
\section{Absorbing-state Potts model}
An important step toward the classification of absorbing state phase 
transitions into universality classes was made by Janssen and Grassberger who
predicted that a large group of models falls into the so-called directed
percolation universality class (DP) and it was conjectured that all
models with a single absorbing state, positive one-component order parameter
and short-range dynamics should generically belong to this
universality class~\cite{DP}.
By now their conjecture received numerous and convincing support.
It also raised a possibility that the number of absorbing states $q$ might be a 
relevant parameter that determines the critical behaviour of a given model.
And indeed, there are some examples that show that models with $q=2$ share the 
same critical behaviour that is called the parity conserving 
(PC) universality class~\cite{PC,HAYE97}.
However, it turns out that PC criticality appears only for models with 
symmetrical absorbing states.
Even a small asymmetry in the dynamics of the model that would favour one of 
the absorbing states over the other will drive the system into the DP 
universality class~\cite{SYMMETRIC}.
Models with $q>2$ were also examined and using a 
relation with some particle systems it was predicted that such models are 
generically in the active phase and only in a limiting case undergo a phase 
transition related with a certain multispecies branching-annihilating random 
walk model~\cite{HAYE97,CARLON}.
As we will discuss in this section, the number $q$ does not fix the critical 
behaviour because some other details of the dynamics play a role too.
For example, we can change the critical behaviour by suppressing certain 
transition rates or extending the range of interactions.
Many factors are therefore responsible for the critical behaviour of a 
given model.

We should mention that absorbing phase transitions also take place in some 
particle systems like e.g., branching-annihilating random walk (BARW).
For such models the critical behaviour is determined, e.g., through some 
symmetries of their dynamics, rather than by the number of absorbing states 
(that is usually the vacuum).
For $d=1$, at the coarse grained level, models with multiple absorbing states 
can 
be related with some particle systems.
For higher dimensions, such an analogy in general does not hold.
There are already some reviews on this subject~\cite{HAYE2000,GEZA2002}.
\subsection{Definition of the model and the Monte Carlo method}
Before defining  our dynamical model, let us recall some basic 
properties of the usual equilibrium Potts model~\cite{WU}.
First, we assign at each lattice site $i$ a $q$-state variable
$\sigma_i=0,1,...,q-1$.
Next, we define the energy of this model through the Hamiltonian:
\begin{equation}
H= -\sum_{i,j} \delta_{\sigma_i\sigma_j},
\label{e1}
\end{equation}
where summation is over pairs of $i$ and $j$ which are usually 
nearest neighbours and $\delta$ is the Kronecker delta function.
This equilibrium  model was studied using
many different analytical and numerical methods and is a rich source 
of information about phase transitions and critical phenomena~\cite{WU}.
To simulate numerically the equilibrium Potts model defined using the 
Hamiltonian~(\ref{e1}), one introduces a stochastic Markov process with 
transition rates chosen in such a way that the asymptotic probability
distribution is the Boltzmann distribution.
One possibility of choosing such rates is the so-called Metropolis 
algorithm~\cite{BINDER}.
In this method one 
looks at the energy difference $\Delta E$ between the final and initial 
configuration and accept the move with probability 
min$\{1,{\rm e}^{-\Delta E/T}\}$, where $T$ is temperature measured 
in units of the interaction constant of the Hamiltonian (\ref{e1}), which was
set to unity.
To obtain a final configuration one selects randomly a site and its state
(one out of $q$ in our case).
In the above described algorithm for $T>0$ there is always a positive
probability of leaving any given configuration (even when the final 
configuration
has a higher energy).
Accordingly, such a model does not have absorbing states for $T>0$.

A nonequilibrium Potts model having $q$ absorbing states can be obtained by 
making the following modification in the Metropolis 
dynamics~\cite{POTTS1,POTTS2}: when all 
neighbours of a given site are in the same state as this site, then this site 
cannot change its state (at least until one of its neighbours is changed).
Let us notice that any of the $q$ ground states of the equilibrium Potts model 
is an absorbing state of the above defined nonequilibrium Potts model.
Moreover, the rules of the dynamics of our model depends on the parameter $T$ 
that for the  equilibrium Potts model would be the temperature.
Although for our model the thermodynamic temperature cannot be defined, we will 
refer to $T$ as temperature.

To study the properties of this model we performed standard Monte Carlo
simulations.  A natural characteristic of models with absorbing states
is the steady-state density of active sites $\rho$.
A given site $i$ is active when at least one of its 
neighbours is in a state different than $i$.
Otherwise the site $i$ is nonactive.
In addition to the steady-state density we also looked at its time
dependence $\rho(t)$.  
In the active phase $\rho(t)$ converges to the
positive value, while at criticality, $\rho(t)$ has a power-law
decay $\rho\sim t^{-\Theta}$.
In addition, we used the so-called dynamic Monte Carlo method where one
sets the system in the absorbing state, locally initiate activity,
and then monitor some stochastic properties of surviving
runs~\cite{GRASSTORRE}.  The most frequently used
characteristics are the survival probability $P(t)$ that the activity
survives at least until time $t$ and the number of active sites
$N(t)$ (averaged over all runs).  At criticality these quantities are 
expected to have power-law decay: $P(t)\sim t^{-\delta}$ and
$N(t)\sim t^{\eta}$.
\subsection{$d=1$}
First, let us consider the case $q=2$.
The simplest possibility is to consider our model on a one-dimensional chain.
However, for $q=2$ and for any temperature $T$, this model is trivially
equivalent to the $T=0$ temperature Ising model with Metropolis
dynamics.  Indeed, in this case the allowed moves are only those which
do not increase energy and they are always accepted.  The same rule
governs the dynamics of the $T=0$ Ising chain.
To overcome this difficulty, we studied our model on a ladder-like lattice, 
where two chains are connected by interchain bonds such that each site has 
three neighbours.

Monte Carlo simulations of the model show that for large enough $T$ the model 
remains in the active (disordered) phase. 
After reducing the temperature below a certain critical value, the model 
collapses on one of the absorbing states. 
The evolution to the absorbing state resembles the coarsening process.
Measuring the critical exponents at the critical point we found that their 
values are very close to those of the PC universality class, which is an 
expected result.

We also did simulations for $q=3$, 4, and 5.
In this case we found that our model remains in the active phase for any $T>0$ 
and collapses on one of its absorbing states only at $T=0$.
It was already suggested that an absence of the transition for models with 
$q>2$ absorbing states is a generic feature~\cite{HAYE97,CARLON}.
Such a conclusion can be obtained relating a model with absorbing states with 
multi-species BARW model that in some case are known to exhibit such a 
behaviour~\cite{TAUBERCARDY}.
However, such a relation is not rigorously established and must be taken with 
care.
And indeed, one can show that in some cases models with $q>2$ absorbing states 
behave differently.

In the following we shall describe a modification  of our nonequilibrium Potts 
model 
that even for $q>2$ undergoes a transition at positive temperature.
To comply with ref.~\cite{POTTS2}, we refer to this modification as a model B.
All we have to do is to introduce the following restriction in the dynamics of 
our model:
a flip into a state different than any of its neighbours is
forbidden.  In other words, we suppress the spontaneous creation of, e.g.,
domains of type A between domains of type B and C.
Here, A, B, and C denote three (out of $q$) different states.
Let us also notice that the above restriction does not break the symmetry
and the absorbing  states of model B are symmetric with respect to its 
dynamics.
Numerical simulations of such a model for $q>2$ show that at positive 
temperature it 
undergoes a phase transition that belongs to the PC universality 
class~\cite{POTTS2}.
It was suggested that with the above restriction, the long-time dynamics of  
model B in the active phase and close to the critical point is dominated by 
parity conserving processes and that might explain the origin of PC 
criticality.
\subsection{$d>1$}
The nonequilibrium Potts model was also studied on higher dimensional lattices 
and below we briefly describe the obtained results~\cite{POTTS1}.\\
(i) $d=2,\ q=2$:\\
Models with a single absorbing state on $d=2$ lattices typically belong to the 
DP universality class.
It is interesting to ask whether for $d=2$ models with double absorbing states 
share the same critical behaviour.
Recent numerical calculations show that indeed there is a group of such models 
that have the same critical behaviour, that was termed the voter universality 
class~\cite{CHATE}.
The name of this universality class was given after a voter model that was 
originally proposed as a model of spreading of an opinion~\cite{LIGGETT}.
Later, various generalizations of this model were also studied~\cite{VOTER}.
In the voter model the order parameter vanishes continuously to 0 upon 
approaching a critical point but the decay is slower than any power law.
In addition, the time decay of the order parameter at criticality is also 
slower than any power law decay and is in fact logarithmic, as it can be shown 
exactly~\cite{KRAP}.
Such an unusual behaviour explains the numerical difficulties in studying 
models of this universality class~\cite{POTTS1}.
The nonequilibrium Potts model for $q=2$ and on  square lattice with nearest 
neighbour interactions also belongs to the voter universality 
class~\cite{POTTS3}.
It was suggested that two-dimensional models with double absorbing state should 
generically belong to the voter universality class~\cite{CHATE}.
However, as we describe below, there are some exceptions from this 
rule~\cite{POTTS3}.

An interesting feature of the voter critical point is the fact that at this  
point actually two phenomena seem to take place.
One of them is the symmetry breaking between two competing states of the model, 
that is similar to the symmetry breaking in the Ising model.
The second phenomenon is the phase transition between active and absorbing 
phases of the model.
It turns out that these two phase transitions can be separated and it happens 
in the $q=2$ Potts model on the square lattice with interactions up to the 
third nearest neighbour.
In this case the behaviour of the model can be thus described as follows.
At sufficiently high temperature $T$ the model remains in the disordered phase.
Upon reducing of temperature, the model first undergoes the symmetry breaking 
phase transition.
Calculation of the Binder cumulant suggests that this transition belongs to the 
Ising type universality class.
Upon further decrease of $T$, the model undergoes a second phase transition 
into an absorbing state.
Since at this point the symmetry is already broken and the absorbing state is 
already preselected, this second transition, as expected, belongs to the DP 
universality class.

The Ising-type phase transition is just one example of a symmetry breaking.
The voter criticality can be regarded as a superposition of this transition 
with DP transition.
One can ask whether other types of symmetries, such as e.g., $Z_3$ or $U(1)$ 
can be superposed with DP.
Possibly in such a case a new critical behaviour might result.\\
(ii) $d=2,\ q=3$\\
In this case (nearest-neighbour interactions) there is a clear evidence of the 
discontinuous phase transition.
In particular, the steady-state density of active sites $\rho$ has a 
discontinuous behaviour
and the time dependent $\rho(t)$ develops a plateau at a critical point.\\
(iii) $d=3,\ q=2$\\
To split the voter critical point in the two-dimensional case we had to include 
further neighbour interactions.
Alternatively, increasing the dimensionality up to $d=3$ also results in two 
separate phase transitions~\cite{POTTS3}.
It would be interesting to check whether in the three-dimensional case some 
additional interactions (possibly antiferromagnetic ones) could actually lead 
to the overlap of these two transitions.

Studying our nonequilibrium Potts model for some values of $q$ and $d$ we were 
tempted to speculate on the overall behaviour of the phase diagram in the 
$(q,d)$ plane~\cite{POTTS1}.
Indeed, it seems that the $(q,d)$ plane is divided into three parts with (i) 
non-mean-field critical behaviour (ii) mean-field critical behaviour, and (iii) 
discontinuous transitions. 
Arrangements of these parts suggests that the qualitative behaviour of our 
model resembles the behaviour of the equilibrium Potts model.
If so, it would imply that the modification of the dynamics that we introduced, 
and that imply the existence of absorbing states, might not change that much 
the qualitative behaviour of the model (as compared to the equilibrium one).

Since our model is formulated in terms of spin-like variables, we can easily 
introduce its various modifications that for example will take into account 
lattice anisotropy, multi-spin interactions, external fields or additional 
symmetries (gauge, $U(1)$,...).
For example some equilibrium homogeneous spin models are known to exhibit 
glassy behaviour~\cite{GLASS}. 
One of the questions is whether a similar behaviour exists when the dynamics 
with absorbing states is used.
\section{Synchronization of dynamical systems}
Recently, synchronization of dynamical, and in particular chaotic, systems 
received considerable attention~\cite{FUJISAKA}.
This is to large extent related with its various experimental realizations in 
lasers, electronic circuits or chemical reactions~\cite{EXP}.
So far, most of the attention has been focused on the behaviour of the 
low-dimensional systems.
More recently, spatially extended, i.e., highly-dimensional, systems are also 
drawing some interest~\cite{EXTENDED}.
Since the synchronized state is an attractor of the dynamics, it can be 
considered as an absorbing state.
Consequently, a transition into a synchronized state (ST) for spatially 
extended systems bears some similarity to absorbing state phase transitions.
However, for continuous dynamical systems, like e.g., coupled map lattices 
(CML)~\cite{KANEKO}, the system cannot reach a perfectly synchronized state in 
a finite time.
This is in contrast with for example some cellular automata that typically can 
reach an absorbing state in finite time.
In some cases such a difference has probably a negligible effect and ST belongs 
to the directed percolation universality class~\cite{LIVI}.
But there are some other arguments suggesting that the critical behaviour at ST 
typically is different and belongs to the bounded Kardar-Parisi-Zhang 
universality class~\cite{PIKKURTHS}.
A possible crossover between these two universality classes is recently 
intensively studied~\cite{LIPDROZSOS,MUNOZPASTOR,AHLERSLIVI}.
\subsection{Coupled-Map Lattices}
To provide a more detailed example we examine a model recently proposed by 
Ahlers and Pikovsky that consists of two coupled CML's~\cite{AHLERS},
\begin{equation}
\begin{pmatrix} u_1(x,t+1) \\ u_2(x ,t+1)\end{pmatrix} =
\begin{pmatrix} 1-\gamma & \gamma \\ \gamma & 1-\gamma\end{pmatrix}\times
\begin{pmatrix} (1+\epsilon\Delta)f(u_1(x,t+1)) \\
(1+\epsilon\Delta)f(u_2(x,t+1))\end{pmatrix},
\label{model}
\end{equation}
where $\Delta v$ is the discrete Laplacian $\Delta v(x)=v(x-1)-2v(x)+v(x+1)$.
Both space ($x$) and time ($t$) are discretized, $x=1,2\ldots,L$ and 
$t=0,1,\ldots$.
Periodic boundary conditions are imposed $u_{1,2}(x+L,t)=u_{1,2}(x,t)$ and,
similarly to previous studies, we set the intrachain coupling $\epsilon=1/3$.
Varying the interchain coupling $\gamma$ allows us to study the transition 
between synchronized (large $\gamma$) and chaotic (small $\gamma$) phases.
Local dynamics is specified through a nonlinear function $f(u)$.

Next, we introduce a synchronization error $w(x,t)=|u_1(x,t)-u_2(x,t)|$ and
its spatial average $w(t)=\frac{1}{L}\sum_{x=1}^L w(x,t)$.
The time average of $w(t)$ in the steady state will be simply denoted as $w$.
In the chaotic phase one has $w>0$, while in the synchronized phase $w=0$.
Moreover, at criticality, i.e. for $\gamma = \gamma_c$, $w(t)$ is expected to 
have a power-law decay to zero $w(t)\sim t^{-\Theta}$.
In the stationary state, and for $\gamma$ approaching the critical value 
$\gamma_c$ one expects that $w \sim (\gamma_c - \gamma)^\beta$.

For $f(u)=2u\ {\rm mod}(1)$, i.e., a Bernoulli map, Pikovsky and 
Ahlers~\cite{AHLERS} found that ST belongs to the DP universality class.
Such a behaviour is in agreement with earlier predictions by Baroni et 
al.~\cite{LIVI} that DP critical behaviour should exist for maps with strong 
nonlinearities (in the case of the Bernoulli map it is even a discontinuity).
Let us notice, however, that DP criticality is typically attributed to models 
with a single absorbing state.
On the other hand, an extended dynamical system as e.g. (\ref{model}) has 
infinitely many synchronized states.
Recently, we applied a dynamical Monte Carlo method to study model 
(\ref{model}).
Our results show~\cite{DROZLIPSYNCH} that exponents $\eta$ and $\delta$ depend 
on the type of a synchronized state, but their sum $\eta+\delta$ remains 
constant.
Such a situation is known to take place in some other models with infinitely 
many absorbing states~\cite{INFINITY}.

A different critical behaviour emerges for the symmetric tent map 
$f(u)=1-2|u-1/2|$.
Since the map is now continuous, as expected, ST belongs to the BKPZ 
universality class.
But what is going on when the symmetricity of the tent map is gradually 
distorted?
In particular, let us examine the following map
\begin{equation}
f(u)=\left\{\begin{array}{ll}
au & {\rm for}~0\leq u<1/a \\
a(1-u)/(a-1) & {\rm for}~1/a\leq u\leq 1, \\
\end{array}
\right.
\label{atent}
\end{equation}
with $1<a\leq 2$.
For $a=2$ this is the symmetric tent map.
Let us notice that in the limit $a\rightarrow 1$, the slope of the second part 
of this map diverges.
In such a limit the map has a strong nonlinearity and we expect that ST in this 
case belongs to the DP universality class.
Numerical calculations show, however, that DP criticality sets in already for 
$a$ in a finite distance from 1~\cite{DROZLIPSYNCH}.
In such a way model (\ref{model}) with the map (\ref{atent}) allows us to study 
the change of the universality class of ST.
An interesting possibility is that at a certain $a=a_c>1$, where DP and BKPZ 
critical lines intersect, a multicritical behaviour appears.
Numerical simulations of model (\ref{model}) are not yet conclusive enough, but 
it is possible that additional insight into this problem can be obtained using 
certain effective models of ST.
This problem is discussed in the next subsection.
\subsection{Interfacial models of synchronization}
An interesting approach to ST was initiated by Pikovsky and 
Kurths~\cite{PIKKURTHS}.
They have argued that the temporal evolution of the small perturbation of a 
synchronized state $w(x,t)$ of the system (\ref{model}) in the continuous limit 
should obey the following Lagevin-type equation~\cite{PIKKURTHS,AHLERS}
\begin{equation}
\frac{\partial w(x,t)}{\partial t} = [a+\xi(x,t)-p|w(x,t)|^2]w(x,t) + 
\epsilon\frac{\partial^2 w(x,t)}{\partial x^2},
\label{langevin}
\end{equation}
where $a$ is a control parameter connected with the transverse Lyapunov 
exponent $\lambda_{\perp}$, that describes an exponential growth of $w(x,t)$.
The Gaussian stochastic process $\xi(x,t)$ has the properties
\begin{equation}
\langle\xi(x,t)\rangle=0,\ \  \langle\xi(x,t)\xi(x',t')\rangle = 
2\sigma^2\delta(x-x')\delta(t-t').
\label{gauss}
\end{equation}
Applying the Hopf-Cole transformation $h={\rm ln}|w|$, Eq.~(\ref{langevin}) is 
transformed into a driven interface model
\begin{equation}
\frac{\partial h(x,t)}{\partial t} = a+\xi(x,t)-p{\rm e}^{2h(x,t)}+ 
\epsilon\frac{\partial^2 h(x,t)}{\partial x^2}+\epsilon[\partial 
h(x,t)/\partial x]^2,
\label{bkpz}
\end{equation}
which is the KPZ equation~\cite{KPZ} with an additional exponential saturation 
term.
In Eq.~(\ref{bkpz}), synchronization corresponds to an interface moving towards 
$-\infty$, and the saturation term prevents the interface from moving towards 
large positive value.
Equation (\ref{bkpz}) is usually referred to as the bounded KPZ equation.

Critical exponents of model (\ref{bkpz})~\cite{TU} remain in a satisfactory 
agreement with those obtained for the CML model (\ref{model}) with the 
symmetric tent map~\cite{AHLERS}.
However, the relation with model (\ref{bkpz}) offers little understanding of 
what changes the universality class into DP for strongly nonlinear local maps 
$f(u)$.
Recently, there were some attempts to understand the emergence of the DP 
universality class.
For example Mu\~noz and Pastor-Satorras~\cite{MUNOZPASTOR}, showed numerically 
that DP might emerge in a more general version of Eq.~(\ref{langevin}).
However, the origin of certain additional terms that appear in their approach 
is not yet clear.

In another approach~\cite{LIPDROZSOS}, we proposed a certain SOS model whose 
dynamics is motivated by the dynamics of synchronization error in CML model 
(\ref{model}).
In this SOS model ST transition is essentially mapped onto a certain wetting 
transition.
It turns out that when certain transition rates, that might be related with a 
binding potential of a wall, decay exponentially fast with the distance from 
the wall $h$, the wetting transition belongs to the DP universality class.
There is a hope that for another form of these transition rates BKPZ 
universality class will be recovered but we still did not succeed to confirm 
it.
On the other hand, when these transition rates decay as a power of $h$, the SOS 
model exhibits a different critical behaviour.
Surprisingly, critical exponents in this case remain in a good agreement with 
those calculated recently for the bosonic version of the PCPD 
model~\cite{CHATEKOCKEL}.
Whether this is just a numerical coincidence or a manifestation of a deeper 
relation is yet to be seen~\cite{COMM1}.
Another approach to examine a change of the universality class in ST using an 
interfacial model was proposed by Ginelli et al~\cite{AHLERSLIVI}.
\section{Conclusions}
In the present paper we described some of our recent results on models with 
absorbing states.
In particular we examined phase transitions and critical behaviour in the 
nonequilibrium Potts model with absorbing states.
We also described the connections between synchronization of spatially extended 
chaotic systems and absorbing phase transitions.
We related the synchronization problem with some models of a driven or wetting 
interfaces.
It would fulfill our goals if this work would stimulate further research in 
this field.
For example, we showed that when the third nearest neighbour interactions are 
included, the voter critical point in the Potts model is splitted.
But this further range interactions are of the same strength as nearest 
neighbour interactions. 
One can consider a model with varying strength of further range interactions.
At a certain amplitude of this interactions the splitting should appear.
At this point three critical lines will meet: voter, Ising and directed 
percolation.
It would be interesting to examine in more details the nature of such a 
tricritical point.
Synchronization transition and its connection with interfacial models is still 
intensively studied by several groups and further interesting results are 
likely to appear.
\acknowledgements{
This work was partially supported by the Swiss National Science Foundation
and the project OFES 00-0578 "COSYC OF SENS".}

\end {document}